# Room-temperature quantum oscillations of static magnetic susceptibility of silicon-carbide epitaxial layers grown on a silicon substrate by the method of the coordinated substitution of atoms

N.T. Bagraev[1,2], S.A. Kukushkin[1 ✉], A.V. Osipov[1], V.V. Romanov[3], L.E. Klyachkin[2],
A.M. Malyarenko[2], N.I. Rul'[1-3]

[1]Institute of Problems of Mechanical Engineering, Bolshoi pr. 61, Vas. Ostrov, St. Petersburg, 199178, Russia

[2]Ioffe Institute, Politechnicheskaya 26, St. Petersburg 194021, Russia

[3]Peter the Great St. Petersburg Polytechnic University, Politechnicheskaya 29, St. Petersburg 195251, Russia

✉ sergey.a.kukushkin@gmail.com

**Abstract.** The article presents the results of measurement and analysis of the field dependences of the static magnetic susceptibility of thin epitaxial silicon carbide films grown on the (110) surface of single-crystal silicon by the method of the coordinated substitution of atoms. In weak magnetic fields, the occurrence of two quantum effects at room temperature was experimentally found: the hysteresis of the static magnetic susceptibility and, in the field dependences, quantum Aharonov-Bohm oscillations of the static magnetic susceptibility. The simultaneous occurrence of these effects is a consequence of two- and one-particle interference of charge carriers (two-dimensional holes) on microdefects consisting of dipole centers with negative correlation energy (negative-U dipole centers).
**Keywords:** silicon carbide on silicon, dilatation dipoles, static magnetic susceptibility, diamagnetism, quantum Aharonov-Bohm oscillations, quantum interference, negative correlation energy, negative-U boron dipoles

*Acknowledgements. The study was supported by the Russian Science Foundation (grant no. 20-12-00193).*

**Citation:** Bagraev NT, Kukushkin SA, Osipov AV, Romanov VV, Klyachkin LE, Malyarenko AM, Rul' NI. Room-temperature quantum oscillations of static magnetic susceptibility of silicon-carbide epitaxial layers grown on a silicon substrate by the method of the coordinated substitution of atoms. *Materials Physics and Mechanics*. 2022;50(1): 66-73. DOI: 10.18149/MPM.5012022_5.

## 1. Introduction

The goal of the presented work is to investigate the behavior of the static magnetic susceptibility of the thin silicon carbide epitaxial layers were grown on single-crystal silicon (110) surface by the method of the coordinated substitution of atoms. The method of the coordinated substitution of atoms is based on the special chemical treatment of the Si surface by carbon monoxide (CO). During this treatment, the following chemical reaction occurs on the Si surface:

2Si (crystal) + CO (gas) = SiC (layer) + SiO (gas) ↑. (1)

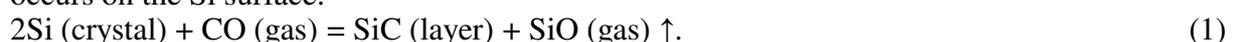

As a result of a reaction (1), the near-surface Si layer transforms into an SiC epitaxial layer. The term "coordinated" means that the new chemical bonds are formed simultaneously and in a coordinated manner with the destruction of the old bonds [1-5]. It was found that the mechanism of the coordinated substitution of atoms preserves the initial silicon cubic lattice, thereby providing the growth of the $3C$-SiC cubic polytype [4,5]. A detailed description of the processes that occur during the SiC growth by the method of the coordinated substitution of atoms, as well as the synthesis technology, can be found in the reviews [1-5].

The main distinguishing property of SiC films synthesized by this method is the formation of an excess concentration of silicon vacancies in them, while in SiC growth by standard methods, there are mainly carbon vacancies. The interaction of silicon vacancies and carbon atoms leads to the formation of the ordered ensembles of carbon-vacancy structures in the SiC layers grown by the method of the coordinated substitution of atoms. The presence of carbon-vacancy structures in SiC gives rise to the unique properties of such films [6-8]. The quantum chemistry and the experimental methods demonstrated that so-called $C_4V$ centers are formed is SiC on Si at synthesis temperature ~1350°C. During the SiC growth by this method on the SiC/Si interface an interfacial layer with a thickness of the order of several nanometers, which demonstrates the extraordinary optical and electric features, appears. It is so due to the process of shrinkage of the initial Si lattice with the parameter of 0.543 nm into the SiC cubic lattice with the parameter of 0.435 nm, which appears during the final stage of the silicon transformation into silicon carbide; this process occurs in the plane of the substrate [1-8]. At the same time, silicon carbide separated from the silicon matrix subjects it to anomalously strong compression, the values of which exceed 100 GPa. At such high pressures, obtaining of SiC with such a good structure would be impossible without the high-accuracy coincidence of every fifth silicon carbide crystal cell with every fourth silicon cell. The shrinkage of the material results in the coordinated arrangement of every fifth SiC chemical bond with every fourth Si bond. The remaining bonds either break, which leads to the vacancies and pores formation, or undergo compression, which leads to the change in the structure of the surface zones of silicon carbide transforming it into "semimetal". For the first time, this phenomenon was discovered by spectral ellipsometry in the photon energy range of 0.5-9.3 eV [6,7]. It was shown that the carbon-vacancy structures result in the appearance of other unique optical, electric, and magnetic features [6-8]. Particularly, the possibility of creating field-effect transistor structures on the SiC/Si layers was demonstrated in [9]. On the basis of the macroscopic quantum properties of obtained structures, the emitters and recorders of the terahertz (THz) frequency range of the electromagnetic spectrum with amplitude and frequency modulation in the entire spectrum were implemented. The possibilities of the spectral dependencies control by using the external characteristics of the sample under study (source-drain current and voltage applied to the horizontal gates in the frameworks of the Hall geometry of the transistor structure) were demonstrated, and the basis of the mechanism of the THz radiation appearance was identified as the quantum analogue of the Faraday's electromagnetic induction [9]. Such an approach opens up huge possibilities for creating coupled systems of GHz–THz frequency range emitter-recorder for the radio-photonics and nano-electronics purposes, in particular, for the possible implementation of quantum-computing operations under the conditions of controlled carriers transport in the obtained structures at high temperatures.

In [8] theoretically using the modeling based on the method of the density functional theory the significant effect of $C_4V$ centers in the SiC/Si on the magnetic properties of silicon carbide was predicted. While the ideal $3C$-SiC grown by the standard methods is a classical semiconductor, $3C$-SiC grown by the method of the coordinated substitution of atoms demonstrates magnetic properties depending on the $C_4V$ concentration ($n_{C_4V}$).

In [8] it is shown that $n_{C_4V}$ can be controlled by varying the time and the temperature of the SiC synthesis.

Thus, the experimental research of the extraordinary magnetic features of the SiC/Si interface grown by the method of the coordinated substitution of atoms was the goal of the presented paper.

## 2. The sample's preparation method

In this study, as the experimental sample, we used the nanoscale SiC layer of ~ (90-100) nm thickness, synthesized on the (100) surface of the single-crystal $n$-type silicon (doped with phosphorus) by the method of the coordinated substitution of atoms [1-5]. After synthesis, the SiC layer was doped with boron under the conditions of non-equilibrium diffusion from the gas phase in the excess flow of the silicon vacancies from the surface of the sample. The synthesis and doping parameters, as well as the sample's characteristics, are presented in Table 1. The formed Hall topology of the transistor structure on the surface of the sample made in the framework of the planar technology has the following parameters: the chip dimensions are $5 \times 5$ mm with 12 field-transistor structures inside the chip.

Table 1. Technological characteristics of the sample

| Substrate | Synthesis conditions | Doping Temperature | Conductivity type, carriers concentration | | Transistor structure |
|---|---|---|---|---|---|
| | | | Before Doping | After Doping | |
| $n$-Si, $20 \, \Omega \cdot cm$, (110) | $T = 1290°C$, $t = 15 \, min$, $P = 2.3 \, Torr$, $I_{CO} = 12 \, sccm$, $\%_{SiH4} = 0.25\%$ | $900°C$ | $p$-type, $\sim 6 \cdot 10^{17} \, cm^{-3}$ | $p$-type, $> 1 \cdot 10^{19} \, cm^{-3}$ | Planar |

As the synthesis conditions here presented: $T$ – synthesis temperature, $t$ – synthesis time, $P$ – CO pressure, $I_{CO}$ – CO flow, $\%_{SiH_4}$ – volume percent of silane (SiH$_4$), which is used during the synthesis [3]. The carriers concentration values were obtained from the capacitive measurements.

## 3. The experimental method

The realization of the "strong-field" criterion ($\omega_c \tau = eB\tau/m^* = \mu B \gg 1$, where $\tau$ – transport time, $m^*$ – carriers effective mass value, $\mu$ – mobility, $\omega_c$ – cyclotron frequency, $B$ – magnetic field value) in the weak magnetic fields at room temperature became possible due to the low values of the carriers' effective mass with their high mobility [7,8]. Thus, the tasks of the study were to measure and analyze the static magnetic susceptibility of the presented sample.

The static magnetic susceptibility was measured by the Faraday method in the magnetic field range of $H = 0 \div 500 \, Oe$ in a step-by-step mode with the 1 $Oe$ step using the automated Faraday Balance setup, based on the MGD 312 FG spectrometer, at room temperature in the regime of the thermodynamic equilibrium sample's state realization. The Faraday method is based on measuring the force acting on the sample with the mass $m$ in an inhomogeneous external magnetic field. The relationship between the static magnetic susceptibility $\chi(T, B)$ and the measured force acting on the sample $F(T, B)$ is determined by the following expression:

$$\chi(T, B) = \frac{F(T, B)}{m \cdot B dB/dz}. \tag{2}$$

The external field gradient $dB/dz$ is provided by the special shape of the pole pieces used in the magnet while the value $B dB/dz$ remains constant throughout the volume occupied by the sample. For the measurements, the sample was placed in a quartz cup connected to the scales by a quartz suspension. The force $F(T, B)$ acting on the sample was defined as the difference between the force of the sample interaction with the field and the force acting on the empty quartz cup, measured under the same external conditions.

The calibration of the experimental setup was carried out using the reference sample – a single-crystal of the magnetically pure indium phosphide with the known value of the static magnetic susceptibility $\chi = -313 \cdot 10^{-9} \, cm^3/g$. The high sensitivity of the MGD 312 FG balance spectrometer in the range of $10^{-10} \div 10^{-9}$ CGS ensures the appropriate calibration stability of $B dB/dz$ values [10].

## 4. Results and discussion

The sample under study reveals a number of features, the analysis of which makes it possible to interpret the results of the experiment, linking them with the structural features of the material, which are obtained due to the original technology of the thin epitaxial SiC layers growth on the near-surface layer of single-crystal silicon.

At low fields, the measured static magnetic susceptibility of the SiC sample grown on the (110) silicon surface demonstrates a transition from the diamagnetic state to the paramagnetic state – dia-para-hysteresis (Fig. 1(a)). Similar behavior of the static magnetic susceptibility was previously discovered in low-temperature experiments on the boron-doped diamond synthesized at high pressures (~ 100 000 $atm$.) and temperatures above 2500 $K$ [11].

As was shown [11, 2], the observed phenomenon is an experimental demonstration of the Meissner-Ochsenfeld effect, which is quite unexpected at room temperature due to the seemingly obvious dominance of the electron-electron interaction in the structure under study. At the same time, the reduction of the electron-electron interaction at room temperature in thin SiC epitaxial layers grown by the method of the coordinated substitution of atoms on the (110) single-crystal silicon surface primarily can arise from the vacancy microdefects presence, which is generated during the synthesis of the SiC layers on the silicon surface, while further boron doping results in the formation of the dipole boron centers with negative correlation energy in the vicinity of the microdefects. The electrostatic field of the negative-U dipole boron centers is exactly responsible for the electron-electron interaction reduction.

As was shown in [9], in the studied structure the "strong-field" criterion at room temperature at an extremely low charge carrier effective mass value is implemented. Moreover, the local phonon mode, which is caused by the dipole boron centers reconstruction, can result in the effective local cooling of the structure at room temperature [13,14].

Within the framework of this approach, apparently, the existence of a superconducting state in the system under study can be accepted, which explains the observed hysteresis of the static magnetic susceptibility of the sample.

Returning to the analysis of the magnetic susceptibility field dependence, it should be noted that the experimental results demonstrate the modulation on the hysteresis curve. The oscillations observed in the weak magnetic field on the background of the hysteresis were interpreted as the macroscopic quantum Aharonov-Bohm effect.

In the presented paper, we only discuss the two mentioned macroscopic quantum magnetic effects, without going into the other features of obtained experimental curve. The Aharonov-Bohm quantum oscillations (A – B) are determined by the change in the value of the magnetic flux passing through the plane of the sample, $\Delta\Phi = \Delta n\Phi_0$, where $\Phi_0$ is the magnetic flux quantum. If the area of an interference circuit $S$ penetrated by the magnetic field is constant, then the magnetic flux quantum is defined as $\Phi_0 = \Delta BS$, and $\Delta B$ is the period of oscillations in the external field. It should be noted, that the experimental observation of the Aharonov-Bohm oscillations on the field dependencies of the static magnetic susceptibility is possible only if the length of the closed carrier trajectory $L$ is less than the phase relaxation length $L_\varphi$. Since in this paper we are talking about the superconducting properties of the low-dimensional structure under study, it should be also noted, that the mentioned condition must be satisfied for both one- and two-particles interference.

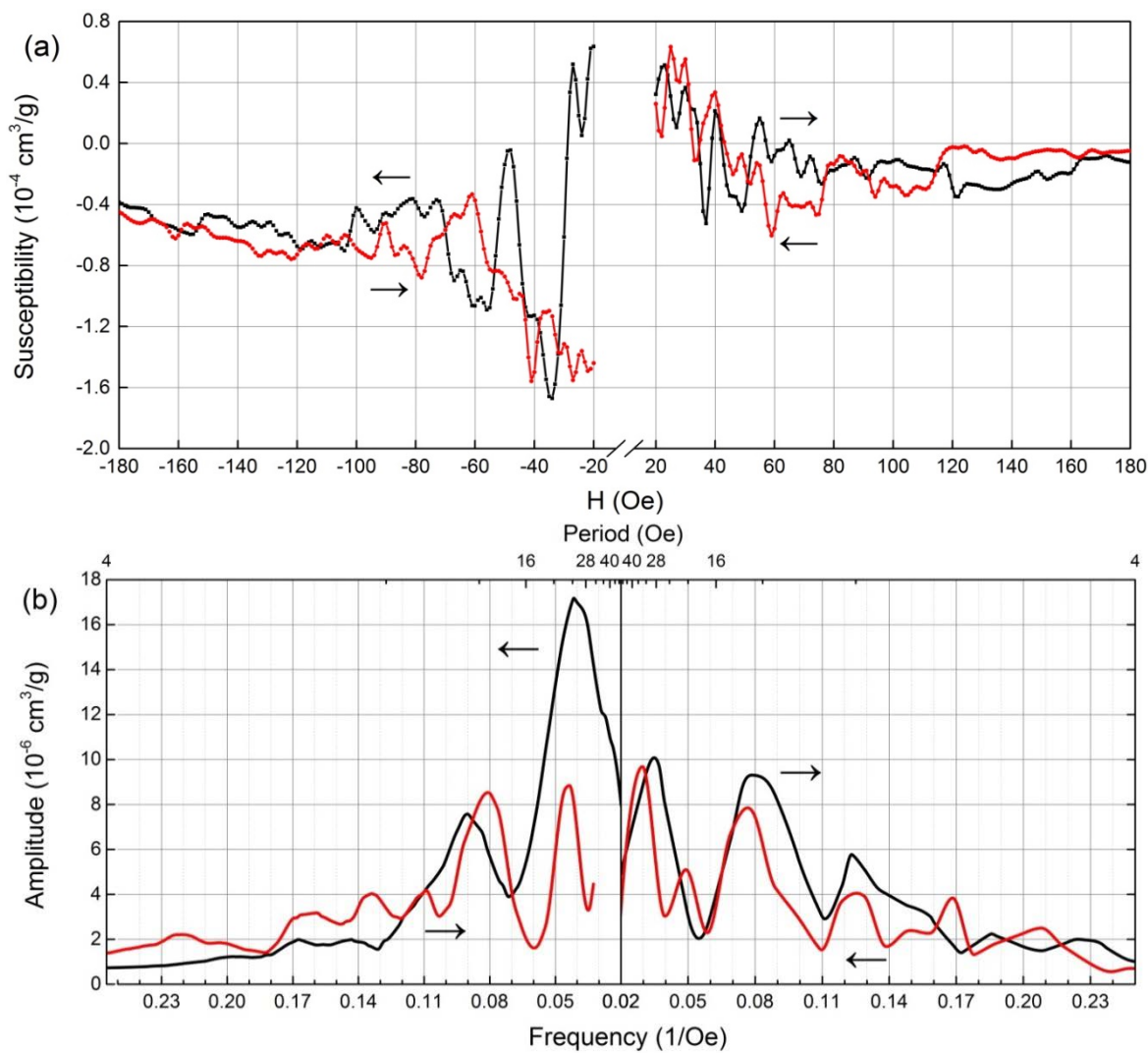

**Fig. 1.** The magnetic susceptibility (a) of the sample for negative and positive directions of the external magnetic field (arrows indicate the direction of increase/decrease in the magnetic field modulus), as well as the Fourier-analysis results (b) for the presented dependencies, respectively

The observation of the A – B oscillations in the SiC nanolayers grown by the method of the coordinated substitution of atoms on the silicon surface determines the condition for the carriers' interference existence on the microdefect of the structure under study and, thus, provides important information about its electric and magnetic properties.

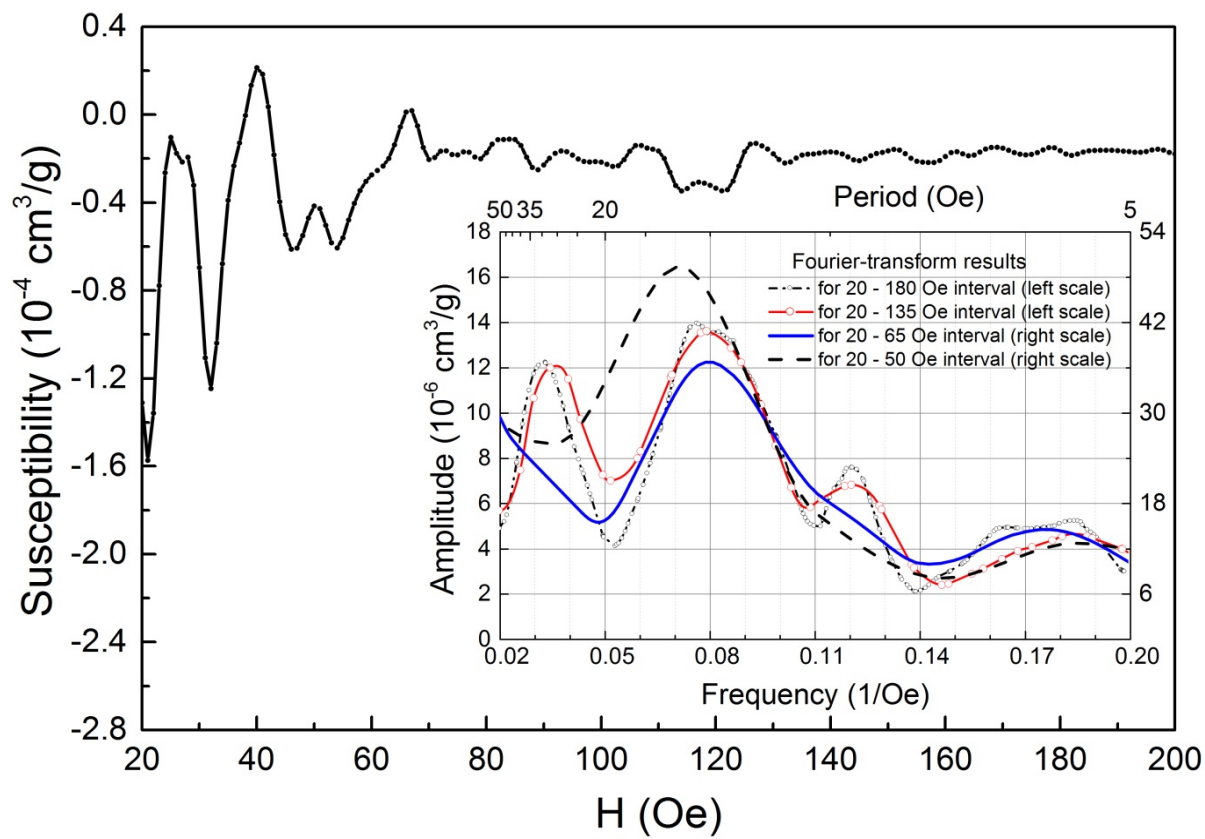

**Fig. 2.** The magnetic susceptibility and the corresponding Fourier spectrum, which demonstrates the existence of two- and one-particle interference of charge carriers in the range of magnetic fields up to 180 *Oe*, as well as the dominance of the superconducting state in the fields range up to 65 *Oe*

The Fourier analysis of the experimental dependencies of the static magnetic susceptibility carried out in this paper made it possible to determine the periods of the A – B oscillations with good accuracy. The oscillations with the periods of 12.5 and 25 *Oe*, which are in our opinion responsible for the appearance of two-particle ($\Phi_0 = h/2e$) and single-particle ($\Phi_0 = h/e$) interference, respectively, and manifest themselves during the process of the single magnetic flux quanta capture, are the most reliably identified in the obtained spectrum.

The occurrence of the above-mentioned oscillations is attributed to the presence of the vacancy microdefects with the sizes of $\sim 1.65 \ \mu m^2$ in the SiC/Si structure under study, which is confirmed both by the corresponding structural research and microdefects similarity to the sizes of the pores formed on the surface of the single-crystal silicon under the SiC layer during the synthesis of SiC.

Here we discuss the effects observed in the range of magnetic fields up to 180 *Oe* (Fig. 2). The analysis of the magnetic susceptibility field dependence of the structure under study demonstrates that in the magnetic field above 50 *Oe* there is a dominance of the A – B oscillation with the period equals to 25 *Oe*, while for the fields below 25 *Oe* the oscillations with the periodicity of 12.5 *Oe* are observed. Thus, it can be assumed that in the vicinity of the magnetic field $H = 50 \ Oe$ there is a critical field value, above which the destruction of the carriers, which is the analogue of the Cooper pair, occurs.

## 5. Conclusions

The presence of the negative-U dipole centers made it possible to realize such qualities as the low effective mass of the carriers and the strong reduction of the electron-electron interaction in the structure under study, which provided the realization of the "strong-field" criterion and determined the possibility of the observation of macroscopic quantum effects in weak magnetic fields at room temperature.

In the SiC structure grown by the method of the coordinated substitution of atoms on the (100) Si surface in weak magnetic fields both one- and two-particle interference were observed, which is possible only when the superconducting state is realized at room temperature.

At low magnetic fields at room temperature, the measured magnetic susceptibility of the SiC sample the transition from diamagnetic state to paramagnetic state – dia-para-hysteresis – was demonstrated, which confirms the occurrence of the high-temperature superconductivity.

## THE AUTHORS

**Bagraev N.T.**
e-mail: Nikolay.Bagraev@gmail.com
ORCID: 0000-0001-8991-6784

**Kukushkin S.A.**
e-mail: sergey.a.kukushkin@gmail.com
ORCID: 0000-0002-2973-8645

**Osipov A.V.**
e-mail: andrey.v.osipov@gmail.com
ORCID: 0000-0002-2911-7806

**Romanov V.V.**
e-mail: romanov@phmf.spbstu.ru
ORCID: 0000-0002-3440-8237

**Klyachkin L.E.**
e-mail: leonid.klyachkin@gmail.com
ORCID: 0000-0001-7577-1262

**Malyarenko A.M.**
e-mail: annamalyarenko@mail.ru
ORCID: 0000-0002-4667-7004

**Rul' N.I.**
e-mail: rul.nickolai@mail.ru
ORCID: 0000-0001-8991-6784